\documentclass[12pt]{article}
\usepackage{a4}

\begin{document}

\begin{titlepage}

\begin{center}
{\Large\bf Lord Rutherford of Nelson, His 
1908 Nobel Prize in Chemistry and Why He Didn't
Get a Second Prize}
\vfill

{\bf C. Jarlskog}\\[0.3cm]

%
%
{Division of Mathematical Physics\\
LTH, Lund University\\
Box 118, S-22100 Lund, Sweden}

\end{center}
\vfill
\begin{abstract}

"I have dealt with many different transformations with various periods of
time, but the quickest that I have met was my own transformation 
in one moment from a physicist to a chemist." 

Ernest Rutherford (Nobel Banquet, 1908)

\vspace{.2cm}
This article is about how Ernest Rutherford (1871-1937) got the
1908 Nobel Prize in Chemistry and why he did not get
a second Prize for his subsequent outstanding 
discoveries in physics, specially the discovery of
the atomic nucleus and the proton. Who were those who
nominated him and who did he nominate for the Nobel Prizes.

In order to put the Prize issue into its proper context,
I will briefly describe Rutherford's whereabouts. 

Rutherford, an exceptionally gifted
scientist who revolutionized chemistry and physics, 
was moulded in the finest classical tradition. 
What were his opinions on some scientific issues such as Einstein's
photon, uncertainty relations and the future prospects for 
atomic energy? What would he have said about the "Theory of Everything"?

\end{abstract}
\vfill
{Extended version of an invited talk presented at the neutrino conference 
"Neutrino 2008", Christchurch, NZ,
25-31 May 2008}
\vfill
\end{titlepage}

\section{Introduction}

I feel as if I am
experiencing "magic" - being here in Christchurch at the
hundredth anniversary of Rutherford's Nobel Prize in
Chemistry. This year, Rutherford is "$1/\alpha$ years old"
and $\alpha$ was indeed his scientific sweetheart.
Christchurch is his university town and the place where he
met his lifelong sweetheart Mary Newton. 
He was very devoted to his
university even though he couldn't visit it so often. New
Zealand is far, very far. It has taken me more than 24 hours
to get here. In his days it was not a question of a day or two
but weeks to get here from Europe! "Within an hour or so
of his death he said to his wife: "I want to leave a hundred
pounds to Nelson College. You can see to it", and again
loudly: "Remember, a hundred to Nelson College."
He hardly spoke after that and on Tuesday evening,
19 October, he died peacefully" \cite{eve}.

I have always been fascinated by Rutherford. He came
from a poor scientific environment and yet rose 
to occupy "the highest position
in the British Empire" \cite{svante}.
A self-made man, and not just a product of
a flourishing environment. An exceptionally impressive
physicist - detector constructer, experimentalist, theorist and a
Nobel Laureate in Chemistry.

\section{"Humble Beginnings"}

Rutherford's birthplace, near the city of Nelson, is
a tourist attraction in New Zealand. There is a small statue there,
of a schoolboy, and an inscription at the site that reads:

"This site is a tribute to one who rose from 
humble beginnings in rural New Zealand to 
world eminence. It is also to show New Zealand 
children that they too can aspire to great heights."
  
This is of course true not only
for children from New Zealand but for all children, provided
there are mechanisms to give them a chance to "rise to
world eminence".

In Rutherford's family there were Mom and Dad plus seven sons
and five daughters. The living conditions were
modest. However "our" Rutherford was exceptionally talented. 
"Although mathematics was his strong subject, he had
no difficulty in obtaining scholarships and prizes for
Latin, French, English literature, history, physics and
chemistry" \cite{eve}. Not only he was good in
calculating (remember his scattering formula) 
but was also excellent
in inventing and constructing apparatuses.
This ability was essential for his success
as an experimentalist. 

It was fortunate that 
Rutherford got a chance, through prizes and scholarships,
to pursue an academic career.
There were scholarships to bring "able young men" to British
Universities and Rutherford was granted such an award.
When his mother came to tell him of his good fortune he was
digging potatoes. He flung away his spade with a laugh,
exclaiming: "That's the last potato I'll dig."

\section{A Rising Star in Cambridge (1895-1898) and His $\alpha$'s }

In October 1895 we find the 24 years old Rutherford 
in Cambridge, England.
He is welcomed to the Cavendish Laboratory by its leader
Joseph John Thomson (1856-1940). Thomson, who later
received the 1906 Nobel Prize in Physics, is generally called
J.J. Here, I shall take the liberty of doing likewise.

\vspace{.1cm}
Rutherford's exceptional talents are quickly recognized
and he becomes somewhat famous in Cambridge. He is invited
to give talks, at several distinguished gatherings, even
at the Royal Society. He demonstrates
his magnetic detector for detection of electrical waves at
(by the standards of the time) large distances.
In his letters to his future wife, he reports on 
his successful presentation:
"No one but myself made any remarks, as it was rather
beyond most of them". His success continues, as he 
notes "My blushing honours are lying thick upon me."

\vspace{.1cm}
Rutherford is much more 
interested in basic science than in industrial applications
 and taking patents. Therefore, 
after the discovery of the R\"ontgen rays, called X-rays
in the English-speaking world, Rutherford changes his
research orientation and starts working with J.J., 
as he explains in a letter to his Mom in July 1896:

\vspace{.2cm}
"I have been working pretty steadily with Professor J. J. Thomson
on the X-rays and find it pretty interesting ...

The method is very simple. A little bulb is exhausted of air
and an electrical discharge sent through. The bulb then lights
up and looks of a greenish colour. The X-rays are given off ...
Aluminium allows the rays to go through easily ..."

\vspace{.2cm}
This is typical of Rutherford. His letters to his Mom and wife
are often full of information on scientific issues.
He often congratulates himself, and, for example, in 1896 
writes to his future wife:

\vspace{.2cm}
"...I am working very hard in the Lab. and have got on
what seems to me a very promising line - very original
needless to say. I have some very big ideas which I hope
to try and these, if successful, would be making of me.
Don't be surprised if you see a cable some morning
that yours truly has discovered half-a-dozen new
elements, for such is the direction my work is taking."

\vspace{.2cm}
If you think Rutherford is bragging too much, 
you should keep in mind that such
comments are a source of enormous joy to a mother
and a sweetheart. Rutherford
often informs his mother of his financial successes
and sends her copies of his diplomas. Again,
a mother who has had to feed many mouths appreciates that. 
In fact his letters to colleagues have a
different style. What the letters have in common is that 
they frequently convey his enormous self-confidence, 
enthusiasm and dedication
to his work, expressed in a custom-tailored language for the receiver.
His letters reveal him as a man with a great deal of social awareness 
as well, far from
the sterotype of the absent-minded professor. He cares about people and 
has an eye for details. He reports, sometimes in detail, about 
how people look or behave, lady's dresses, etc.

\vspace{.1cm}
What Rutherford discovers in Cambridge is 
that X-rays make ions. By charge separation, he discovers 
that there are
two kinds of rays, which he calls alpha rays and beta rays,
and embarks on closer studies of their properties,
especially the less known and more energetic alpha rays.

\vspace{.1cm}
Getting a good permanent job in academia is difficult 
even for a person of Rutherford's calibre.
Rutherford, as he writes in several of his letters, 
is not interested in going to a science's barren soil
and start building research facilities. He
is a man of action working in an exploding new area
of research - radioactivity - a field of international
knife-sharp competition that attracts leading
scientist. He can't afford to waste any time. 

\section{At McGill (1898-1907) and the Transmutation of Elements}

Late in 1898, Rutherford, at the age of 27, becomes 
Macdonald professor of physics at McGill University,
Montreal, Canada.
He considers himself, as he states in
a letter to his Mom, 

\vspace{.2cm}
"extraordinarily lucky to start
so well, for not one man in 10,000 ever gets the
opportunity ...". 

\vspace{.2cm}
The working conditions at McGill are fine,
thanks to very generous donations by a philanthropist, 
Sir William C. Macdonald (1831 - 1917), who years later remarked
that all his expenditure was fully justified by
Rutherford's results alone.

\vspace{.1cm}
At McGill, Rutherford is the successor of a famous man,
who had decided to return to England and
who according to Rutherford, "was considered a
universal genius" by the locals. (I have seen that
this gentleman later 
received three nomination to Nobel Prize in Physics.) 
There is a general
unhappiness about his departure. When a scientist
laments about how sorry they all are to have lost him J.J.
declares "I don't see why you should be, you
got a better man anyway." 

\vspace{.1cm}
Rutherford's work at McGill is outstanding.   
His sensational finding is that
atoms are not necessarily eternal, they 
can transform into one another:
transmutation of elements. He proposes the 
"genealogical tree" of the uranium family where 
he even has to postulate the existence of
a yet unseen intermediate state in the chain.
This is no less than a revolutionary idea.  

\vspace{.1cm}
The great authority of the time William Thomson (Lord Kelvin, 1824-1907) 
and his co-writer Peter Tait
had reported that:

\vspace{.2cm} 
"the inhabitants of the earth cannot continue
to enjoy the light and heat essential to their 
life for many million  years longer, unless sources now 
unknown to us are prepared in the great storehouse of creation".

\vspace{.2cm}
Rutherford applies his findings in radioactivity
and discovers that the sun will shine much longer.
He writes to his wife "My attendance keeps up steadily
and all sorts of people turn up to hear them. In my
lecture tomorrow I am expecting a large audience as
I am dealing with questions
of the effect of radioactivity on the age of the sun
and earth. ... I have had a round of visits during the week - dinner,
lunches, teas. ..." 

\vspace{.1cm}
He is now a Celebrity. He makes headlines in the newspapers,
such as "Doomsday postponed".
There is a great deal of 
demand on his time. He is frequently the guest of
honour at important events, gets prizes and medals,
is elected into distinguished societies, such as the
Royal Society, 
and is offered "a year's salary for ten lectures"
to be given at Yale and so on. He makes his Mom happy by,
writing to her:

\vspace{.2cm}
"If you get the August number of Harper Magazine you will see
a photo of my noble self...."

\vspace{.2cm}
Having been informed that the 1904 Nobel Prizes in Physics and
Chemistry have been awarded to John W. Strutt (Lord Rayleigh) and 
William Ramsay, he writes in a letter to his wife: 

\vspace{.2cm}
"I think they are a very good selection.
I may have a chance if I keep going, in another ten years,
as there are a good many prominent physicist like J.J. and 
others to have their turn of spending the money. It is just as well too
that I have got something worth having to look forward to ..."

\vspace{.2cm}
Rutherford hardly ever made any errors. Here he makes two
at one swish! It 
took only four years and not ten years, and it wasn't physics
but chemistry.  

\section{Nominations to Nobel Prize in Physics and Chemistry}

In order to be eligible for a Nobel Prize in physics
or chemistry, the candidate must have been nominated
for the year in question. All that is required is just one valid
nomination, i.e., a nomination by a person who has been
invited to nominate.

Rutherford is nominated to the Prize in 1907 and 1908.
In 1907 he has seven nominations to the Physics Prize 
and one to the Chemistry Prize. In 1908 he receives five
nominations in Physics and three in Chemistry. This amounts to
a total of 16 nominations for the two years 1907-1908.
13 of these 16 nominations come from Germany, two from Sweden and
one from Canada. 
His nominators for the Physics Prize in 1907 are:

\vspace{.2cm}
$\circ$ Adolf von Baeyer (1905 NLC)
 
$\circ$ Hermann Ebert 

$\circ$ Vincenz Czerny 

$\circ$ Emil Fischer (1902 NLC)
 
$\circ$ Philipp Lenard (1905 NLP)
 
$\circ$ Max Planck (later 1918 NLP) 

$\circ$ Emil Warburg 

\vspace{.2cm}
\noindent Here NLC/NLP stand for Nobel Laureate in Chemistry/Physics. 
All of these nominations 
come from Germany and are written in German. His nominator
to the Chemistry Prize is Svante Arrhenius, Sweden
(see appendix B at the end of this article).
In 1908, his nominators to Physics Prize are:

\vspace{.2cm}
\hspace{1cm} $\circ$ Arrhenius, $\circ$ John Cox, $\circ$ Lenard, 
$\circ$ Planck, $\circ$ Warburg.

\vspace{.1cm} \noindent
where the only "newcomer", Cox, is a professor at McGill.

\vspace{.2cm}
\noindent He is nominated to the 1908 Chemistry Prize by  

\vspace{.2cm}
\hspace{1cm} $\circ$ Arrhenius, $\circ$ Oskar Widman, $\circ$ Rudolf Wegscheider

\vspace{.2cm}
\noindent the first two from Sweden and the latter from Austria.

\vspace{.1cm}
Most of the above nominations 
are short letters of a few lines. Some of the 
nominators attach a list of references but others
take it for granted that Stockholm knows Rutherford's
work. They state that he deserves 
the Prize for his work on radioactivity.
Let me give a few examples. Lenard writes 
that he would like to nominate
Rutherford for his work on radioactivity and (my 
translation):

\vspace{.1cm}
"specially for the first proof of the chemical
transformation of an element/the radium/ Phil. Mag.
Nov. 1904."

\vspace{.2cm} \noindent
Planck nominates him for his experiments and
research on radioactivity and adds (my translation)
 
\vspace{.1cm}
"for having to some extent swept away
the blanket of darkness that still enwraps the
nature of these processes."

\vspace{.1cm} \noindent
Wegscheider, from Vienna, writes (my translation):

\vspace{.2cm}
"This Rutherfordian idea is of such importance to chemistry
that I have no problem with recommending him to 
the Chemistry Prize even though he is a physicist."

\vspace{.2cm}
\noindent The longest nomination letter comes from John
Cox at McGill, and
is dated February 8, 1907. The letter arrives
after the deadline (January 31, 1907) and, therefore, is not 
valid for 1907 but is saved as a
nomination for 1908. Cox writes:

\vspace{.2cm}
"Gentlemen,
 
 \vspace{.1cm}  
   In response to the invitation which I had the honour
   to receive from you to propose a candidate for the
   Nobel Prize in Physics for the year 1907, I beg leave
   to suggest the name of my colleague Professor
   Ernest Rutherford, one of the Macdonald Professors
   of Physics in McGill University.
   
   \vspace{.1cm}
   I regret that I did not observe with sufficient
   care at the time that such proposals should be
   made before February 1st. But it is almost certain
   that Professor Rutherford's name will have been
   brought before you from other quarters; so that
   I deem it well to forward herewith a partial list
   of his scientific work, which may help to support
   other proposals even if this one should be excluded
   by the date. 
   
   \vspace{.1cm}
   Professor Rutherford is leaving us in the autumn to
   occupy the chair of Physics in the Victoria University,
   Manchester, England. It would indeed be a satisfaction to
   his friends here, if he should receive so great an
   honour while still a member of the University where during
   nine years he has completed so many researches.
   
   \vspace{.1cm}
   I have the honour to be, Gentlemen, with the highest
   respect,
   
   \vspace{.1cm}
   Obediently Yours
   
   John Cox
   
   Macdonald Professor of Physics and Director of
   the Macdonald Physics Building."
   
   \vspace{.1cm}
   John Cox, who is 20 years older than Rutherford, had been
   one of the two "head-hunters" who had 
   interviewed Rutherford for the professorship at McGill.
   Eve tells us that Cox, before arrival of Rutherford,
   had remarked to him that 
   he was feeling rather dispirited because there seemed
   nothing new going on in Physics. The main things,
   he said, had all been
   found out and the work which remained was to carry on a
   great number of experiments and researches into 
   relatively minor matters. 
   When Rutherford got going, Cox was ready and glad to sing another
   tune. Cox is now a loyal supporter of Rutherford.
 The scientists at McGill are 
 worried that Rutherford's
 revolutionary ideas about the transmutation of elements
 might turn out to be wrong and bring discredit on the University.
 Cox, the Director of Physics, rises to 
 defend Rutherford and predicts that
 "some day Rutherford's experimental work would be
 rated as the greatest since Faraday ..." \cite{eve}
   
\vspace{.2cm} \noindent
Returning to the nominations to the Chemistry Prize, the
one by Oskar Widman differs from the others as he proposes
that Rutherford should share the Prize with his former research
student (postdoc in modern terminology),
Fredrick Soddy, while all the other nominators opt
for an undivided award to Rutherford.
(Widman was a member of the Nobel Committee for
Chemistry during 1900-1928. We will meet him again
in the next section.)
    
 \vspace{.1cm}
 You may wonder about J.J., who had always been 
 very supportive of Rutherford. Why doesn't he
 nominate his great student? Actually, he does, by
 submitting a nomination in 1908, which, 
 however, arrives too late
 and is therefore invalid for that year but
 is saved for the 1909 Prize. By
 then, however, Rutherford has received the 1908 Prize
 thus making J.J.'s nomination invalid! The rules did not allow
 the nomination of a person who had received the Prize
 within the previous two years! Thus Rutherford had no
 nominations from England or France, where his work
 was very well known and where there were 
 qualified nominators, among them several
 Nobel Laureates. 
 
 \section{Deliberations on Nobel Prize to Rutherford}
 
 Rutherford is nominated for his work on radioactivity, the
 essential issue being the decay of radium. The
 Nobel Committee for Physics, in its 1907 report to the Academy,
 brushes him aside quickly by stating: 
 
 \vspace{.2cm}
 "..his observation of the decay of a chemical element
 (radium) should be awarded with the Chemistry Prize
 rather than the Physics Nobel Prize. Therefore, we deem
 we should not suggest him as a recipient of this year's
 Nobel Prize in Physics."
 
 \vspace{.2cm} \noindent
 In other words, radium is
 a chemical element and that's chemistry. 
 This matter is not trivial. The 1904 Nobel Prizes
 in Physics and Chemistry were awarded respectively to John William
 Strutt (Lord Rayleigh) and William Ramsay.
 Both of them received the Prize for the discovery of chemical elements.
 Strutt was a physicist and Ramsay a physical chemist.
  
 \vspace{.1cm}
 The Nobel Committee for Chemistry, in its 1907 report to the Academy,
 states that: 
 
 \vspace{.2cm}
 "Rutherford has been nominated for his studies of radioactivity,
 by seven nominators to the Physics Prize and by one
 nominator to the Chemistry Prize. This is understandable,
 taking into account that Rutherford uses 
 physical methods while
 the results, so far as they are concerned with
 chemical elements, must be considered to be of fundamental
 importance also for chemistry."
 
 \vspace{.2cm}
 \noindent The Committee then opts for a wait-and-see strategy.
 
 \vspace{.1cm}
 In 1908 the Nobel Committees for Physics and Chemistry meet
 and decide that Rutherford's work is more relevant to chemistry
 than to physics. Svante Arrhenius (see Appendix A) is worried
 that Rutherford might actually fall between
 two stools at the Academy's plenum, where the final decision
 is made. There the objection could be raised that he is not a 
 chemist and the physicists
 have already opted for someone else. He writes to the Academy
 proposing that "if the Academy should decide that it
 is not appropriate to give him the Chemistry Prize he
 should be awarded the 1908 Physics Prize".
 
 \vspace{.1cm}
 Contrary to the Physics Committee, the Chemistry Committee 
 takes its candidate Rutherford very seriously.
 Their report to the Academy contains about 15 pages on him!
 Here Rutherford's competitors are the almost 40
 years older Sir William Crookes
(1832-1919) and to a lesser extent Rutherford's
former research student Fredrick Soddy (1877-1956). 

\vspace{.1cm}
Crookes, who is nominated, primarily for his life-work, by 
William Ramsay and Silvanus Phillips Thompson,
is a remarkable scientist. He 
has discovered thallium in 1861 and electrons (cathode rays)
in the second half of 1870's, using
discharge tubes. The Committee preferes recent discoveries.
Therefore, Crookes' recent discovery  
of uranium-X in 1900 is considered to be the most relevant.
 
\vspace{.1cm}
Soddy has worked with Rutherford at McGill, 1900-1902,
and has written a number of seminal papers with him. 
As mentioned before, he has been nominated "internally", 
by the Committee member
Widman, to share the Prize with Rutherford. 
 
 \vspace{.1cm}
 The central issue, in the case of 1908 Prize, is "emanation", 
 i.e., a chemical element giving birth to something else. 
 The Chemistry Committee's report on Rutherford is much
 too long to be reproduced in this paper. Therefore, 
 I will only give a few excerpts from it, 
 marked by bullets here below. The Committee says:  
 (again my translation):
 
 $\bullet$ Rutherford's publication of discovery of thorium 
 emanation predates that
 of Crookes' discovery of uranium-X.
 
 (Here, I should add that this statement is not quite correct.)
 
 $\bullet$ Rutherford has done both experimental and theoretical work.
 
 $\bullet$ His experimental work concerns the study of transformation
 of radioactive elements and the genetic relationships
 between them.
 
 $\bullet$ Rutherford's theoretical work contains the formulation and 
 development of the so called decay hypothesis, for describing
 the transformation of elements and deducing the laws that
 govern them.
 
 $\bullet$ Rutherford has found the most exact method to compare the 
 intensity of the emitted radiation whereby radiation 
 phenomena can be studied quantitatively.
 
 $\bullet$ It is due to his work in 1899 that the 
 radiation emitted by radioactive
 materials could be classified into the categories which
 he has called alpha rays and beta rays. In addition, he has
 called the radiation discovered in the following year, by
 Villard, gamma rays.
 
 $\bullet$ Alpha rays were less known than beta rays until
 Rutherford discovered their vital role in
 radioactive phenomena and found that they carry the largest portion
 of the emitted energy in the form of ionising rays,
 much more than beta rays.
 
 $\bullet$ He has shown that alpha rays are deflected by both electric and
 magnetic fields, and that they are positively charged. He
 has proposed that alphas are doubly charged helium
 atoms.
 
 $\bullet$ Rutherford has insisted on the material nature of the
 emanation process and has done experiments to verify
 his hypothesis.
 
 \vspace{.2cm}
 Here I would like to add a short aside, as the latter 
 point was a matter of much dispute.
 As an example, one of the greatest authorities of the time, 
 William Thomson
(Lord Kelvin) declared that radium receives its energy by
absorption of ethereal waves. You may have read in your textbooks
that Michelson and Morley had discovered, already in 1887,
that there is no ether. That is a widely propagated misconception. 
Ether was alive until a much later date, but that is not
a subject that will concern us here. I would only like
to quote what Rutherford said about ether:

\vspace{.2cm}
"With regard to the question "What is Electricity?" so often
asked the scientist by the layman, science cannot at present
venture an adequate answer. ....Attempts have been made to
explain electricity as a manifestation of the universal
medium or ether ...Even if we may ultimately explain
electricity in terms of ether, there remains the still
more fundamental problem, "What is ether?". An attempt
to explain such fundamental conceptions seems of necessity
to end in metaphysical subtleties."

\vspace{.2cm}
These words were uttered in 1905, almost two decades after the
Michelson-Morley experiment, and by one of the greatest
scientists of the time.

\vspace{.1cm}
Returning to radioactivity, some other leading 
scientists attributed the emanation to some kind of
"storable energy". The state of affairs was highly confused
and it took Rutherford's genius to sort it out.
 
The Chemistry Committee's report continues on and on 
about Rutherford's
 ingenious experiments and his deep insight regarding
 what was going on in the complicated chain of the
 emanation processes. Here are some more extracts: 
 
 $\bullet$ Rutherford had even predicted
 the existence of a not yet observed intermediary state in order
 to get a comprehensive description of the emanation chain. 
 
 $\bullet$ Research on radioactivity had, in just a few
 years, led to 
 a large number of surprising observations,
 which appeared to be incompatible with previously undisputed 
 doctrines. The mysterious transformation of a
 chemical element into another one 
 appeared to be contrary to chemistry's
 underlying hypothesis of immutability of elements. 
 And where did the enormous energy released in
 these processes come from?
 
 $\bullet$ Rutherford had thus shaken the foundations of chemistry
 by replacing its assumption of the immutability
 of chemical elements with a new and more general hypothesis.
 
 The report describes the theory of 
 Rutherford and Soddy, their introduction of the 
 exponential decay law, lifetimes, etc. 
 
 $\bullet$ The theory had gained ground quickly. The half-lives
 were found to vary within a large range, 
 from a few seconds to several thousand million years.
 
 $\bullet$ Although
 the details were not yet well known, there could be no
 doubt that the disintegration hypothesis had been
 an exceptionally fruitful working hypothesis.
 
 $\bullet$ 
 Evidently, the acquired
 knowledge about radioactivity is not the work of just one
 person. Many people have been involved. However,
 Rutherford's share is of such importance
 that he is an undisputable leader in this field. More so
 as the time has gone by and his followers have confirmed
 the results of his research. He deserves the Nobel
 Prize in Chemistry without a shadow of doubt.
  
 $\bullet$
 A more difficult question concerns whether any of
 Rutherford's collaborators should share the Prize with him.
 
 $\bullet$ He has had a large staff of assistants and 
 about half of
 his articles on radioactivity have been signed by two authors.
 However, a closer study of his work shows that most of
 his assistants had helped him with limited particular tasks
 and that their contribution has been secondary as compared to
 Rutherford's. The only exception is the case of Soddy, who
 not only was a collaborator of his, on some of his most
 important experimental studies 1902-1903, but also
 participated in the formulation of the theory of disintegration
 of elements. Naturally, the question of their individual 
 contributions, in formulating this theory, 
 can not be accessed by outsiders.
 It is remarkable that none of the foreign nominators 
 have suggested that Soddy should share the Prize with 
 Rutherford.
 
 \vspace{.1cm}
 Finally, the Committee argues against honouring Soddy,
 together with Rutherford because:
 
 $\bullet$
 a shared Prize could easily be misinterpreted 
 as an underestimation of the eminent importance
 of Rutherford's work for chemistry and more generally for
 modern natural sciences, specially since the Chemistry Prize,
 up to now, 
 has only been awarded to one laureate at a time.
 
 \vspace{.2cm}
 In fact this tradition was kept
 until 1929, when for the first time the Chemistry Prize was
 jointly awarded to two people. In contradistinction, in physics,
 from an early date, there were joint and divided awards. 
 The first one was given already in 1902 to Lorentz and Zeeman.
 
 \vspace{.1cm}
 What about Sir William Crookes? The Chairman of the Chemistry
 Committee, Otto Pettersson, writes a report to the Academy
 proposing that the
 1908 Prize be awarded to Crookes on the gounds of his
 seniority and that he deserves the Prize.
 The 37 year old
 Rutherford could wait a little and yield the Nobel rights to 
 the 76 year old Crookes. Finally, however, Pettersson
 decides to support the decision of the other members
 of the Committee on the grounds of "the exceptional importance   
 of Rutherford's discoveries". 
 
 \vspace{.1cm}
 Rutherford "eclipses" his competitors. He is judged to be 
 an epoch-maker, a solid, precise scientist and an undisputed 
 leader. He does systematical
 work, carried all the way to completion and draws (theoretical) 
 conclusions out of the results. His work has had a huge impact on 
 the progress of science.
    
 \vspace{.1cm}
 We don't know what
 went on at the Academy when the case of Rutherford was
 brought up by the physicists and chemists. No
 minutes are taken on such occasions. The outcome was what we all know:
 Rutherford is awarded the 1908 Nobel Prize in Chemistry:

\vspace{.2cm} 
"for his investigations into the disintegration of the elements,
and the chemistry of radioactive substances".

\vspace{.2cm}
Rutherford's Nobel Lecture can be found in \cite{ruth}.
He talked primarily about his $\alpha$ particles.
As I quoted in the Abstract of this paper, in his 
short after dinner speech at the Nobel banquet Rutherford had said:

\vspace{.2cm}
"I have dealt with many different transformations with various periods of
time, but the quickest that I have met was my own transformation 
in one moment from a physicist to a chemist." 

\vspace{.2cm}
In a letter dated 24 December 1908, Rutherford writes to his Mom:

\vspace{.2cm}
"I am sure that you have all been very excited to hear
that the Nobel Prize in Chemistry has fallen my way.
It is very acceptable both as regards honour and cash. ...
We have just returned from our journey to Stockholm,
where we had a great time - in fact, the time of our
lives."

\section{Rutherford the Nominator}

As a Nobel Laureate, Rutherford was automatically
invited to nominate Nobel Prize candidates. His nominees
in physics were

\vspace{.2cm}
$\circ$ 1912: John H. Poynting

$\circ$ 1918: Charles G. Barkla (1917 NLP)

$\circ$ 1919/22: Niels Bohr (1922 NLP)

$\circ$ 1924/26/27: Charles T. R. Wilson (1927 NLP)

$\circ$ 1929: Owen W. Richardson (1929 NLP)

$\circ$ 1930: Chandrasekhara V. Raman (1930 NLP)

$\circ$ 1935: James Chadwick (1935 NLP)

$\circ$ 1937: John D. Cockroft and Ernest T. S. Walton (both 1951 NLP)

\vspace{.2cm}
Here NLP stands for Nobel Laureate in Physics. We see that 
Rutherford did extremely well in suggesting suitable candidates.
Except Poynting, they all got the Prize and most of them
in the very same year that Rutherford nominated them
for the first and thus the only time.
It appears as if Barkla got it even the year before!
In fact Rutherford's
nomination of Barkla was the only nomination Barkla
ever had. As this nomination is particularly interesting
I will return to it later.

Concerning Poynting, Rutherford nominated him for his
contributions to experimental and theoretical physics,
gravitation, the pressure of light and transfer of 
energy in the electromagnetic field. Information on
his other nominees is easily found, for example by going
to the internet site http://nobelprize.org/.

\vspace{.1cm}
A great scientist who Rutherford surely would have nominated
was Henry Moseley (1887-1915), who had worked with him in 
Manchester. Rutherford thought very highly of him.
Moseley was nominated to both
Physics and Chemistry Prizes in 1915 by Arrhenius. Unfortunately,
later that year he was killed in the war.

\vspace{.1cm}
Rutherford's nominees in chemistry were: 

\vspace{.2cm}
$\circ$ 1912: William Henry Perkin, Jr.

$\circ$ 1918/1919/1922 Fredrick Soddy (1921 NLC)

$\circ$ 1935: Fredrick Joliot and Irene Joliot Curie (both 1935 NLC)

\vspace{.2cm}
Here, again, NLC stands for Nobel Laureate in Chemistry.
In 1922, Soddy was awarded the 1921 Nobel Prize
in Chemistry. In addition to the above three nominations
from Rutherford and the one by Widman, that I 
discussed above, Soddy had only one more nomination 
(from Wilhelm Schlenk in 1918). 

\section{Manchester Period (1907 - 1919) - the Discovery of the 
Nucleus and the Proton}

Already in 1901, Rutherford writes from McGill to J.J.:

\vspace{.2cm}
"I think you know fairly well my position here.
The laboratory is everything that can be desired, ...
(I) greatly miss the opportunities of meeting men
interested in physics.  ...I think that this feeling of
isolation is the great drawback to colonial appointments,
for unless one is prepared to stagnate, one feels
badly the want of scientific intercourse." 

\vspace{.2cm}
So when
the opportunity arises, for a professorship
in Manchester, Rutherford takes it. Here, he is, in his own words,
very fortunate 
to find a most competent assistant, Johannes (Hans) Geiger 
(1882 -1945) whom he praises in several of his letters:

\vspace{.2cm} 
"He is a very
excellent experimenter and is a great assistance to me"

\vspace{.2cm}
"I have never worked so hard in my life ...
Geiger is a good man and worked like a slave".

\vspace{.2cm}
Rutherford and Geiger succeed in counting alpha 
particles one by one,
thus enabling Rutherford to determine their charge.
Geiger would fire alpha particles through thin metal
foils and measure their, what we would call, scattering
angle. 
In a letter in 1911, to Otto Hahn
(with whom Rutherford corresponded frequently)
Rutherford writes

\vspace{.2cm}
"I have been working recently on scattering of alpha and
beta particles and have devised a new atom to explain the
results, and also a special theory of scattering. Geiger
is examining this experimentally, and finds so far it is
in good agreement with the facts. I am publishing
a paper on the subject to appear shortly."

\vspace{.2cm}
This alludes to the famous Rutherford model of the atom,
with a compact nucleus inside, and to his scattering formula.

\vspace{.1cm}
During his Manchester period, Rutherford makes another
striking discovery. On bombarding nitrogen with his beloved alpha
particles he discovers a new particle which he calls the proton. 
He publishes this
just before leaving Manchester in 1919.

\section{Return to Cambridge, 1919 - 1937}

In 1919, Rutherford returns to Cambridge. He, a
Super-Celebrity, has been appointed to succeed J.J.
as the Director of the Cavendish Laboratory.
He continues his work on protons, by shooting alpha
particles at light atoms. His technical assistant,
G. R. Crowe, has witnessed on Rutherford's
active engagement in the experiments
and his sense of humour. Rutherford
checks Crowe's set up by asking several questions,
did you do this or that, and declares:  

\vspace{.2cm}
"Crowe, my boy, you're always wrong until I've proved
you right! Now we'll find their range."

\vspace{.2cm}
\noindent This concerns the range of protons, which were knocked out 
of the light atoms.

\vspace{.1cm}
\noindent Rutherford predicts the existence
of the neutron, deuteron, tritium and helium-three. In 1921
he sets out to discover the neutron but doesn't succeed. 

Further honours are bestowed on Rutherford and he 
makes a transition from   
a Super-Celebrity to a Hyper-Celebrity.
Nonetheless, he writes papers with his fellow researchers
and makes further discoveries until the end of his life.

\section{A Second Prize to Rutherford?}

Usually Nobel Laureates are not nominated for
a second Prize, though there are some exceptions. 
Einstein, for example, was never nominated after
1922, the year in which he received the 1921 Prize. 

\vspace{.1cm}
Rutherford had received the 1908 Prize in Chemistry
and subsequently had made stunning discoveries in
physics. So, one might have expected that he would be
nominated to the Physics Prize. After all
Marie Curie had been awarded both Prizes. 

\vspace{.1cm}
The archives reveal that Rutherford was actually
nominated for a second Prize, a Prize in Physics,
but only by three people. These were:

\vspace{.2cm}
$\circ$ The (Theodor) Svedberg: 1922/1923

\vspace{.2cm}
$\circ$ David S. Jordan: 1924

\vspace{.2cm}
$\circ$ Johannes Stark: 1931/1932/1933/1935/1937. 

\vspace{.2cm}
Just for completeness, I should add that he also received
a nomination to a second Prize in Chemistry. That
came from the 1911 Nobel Laureate in Physics, Wilhelm Wien. 
This nomination was marked as invalid on the grounds that 
the discoveries, for which he was nominated, 
fell outside the realm of chemistry.   

\vspace{.1cm}
His first nominator, The (Theodor) Svedberg is
a distinguished member of the Academy
(see appendix B at the end of this article).
He nominates Rutherford in 1922 
for his atomic model. The hottest candidates that 
year are Einstein and Bohr, who have  
respectively 17 and 11 nominations. Svedberg wants 
Rutherford to be awarded the Physics Prize before Bohr,
his argument being that Bohr is nominated for his atomic
model which is based on Rutherford's model.

\vspace{.1cm}
The Committee, in its 1922 report to the Academy, argues 
against Svedberg's proposal, on the grounds that:

\vspace{.2cm}
"giving Rutherford a Prize in Physics would imply
that the 1908 decision to award him the Prize in Chemistry was
wrong because the methods used in these discoveries are similar
and the Bohr model of the atom is superior to Rutherford's".

\vspace{.2cm}
As an aside, I would like to mention that 
Niels Bohr (1885-1962) had gone to Cambridge in 1911 
to do experimental work 
with J.J. but had left the year after to work with
Rutherford in Manchester. 
One of the 11 nominations of Bohr came from
Rutherford. The letter shows Rutherford's great appreciation of
his former research fellow. 

\vspace{.1cm}
The outcome in 1922 is that Bohr gets the  
1922 Nobel Prize in Physics and Einstein the 1921
Prize which had not yet been awarded.

\vspace{.1cm}
In 1923, Svedberg repeats his nomination, adding another superb
discovery of Rutherford: the proton. This means that 
the matter has to be considered more seriously. Svante Arrhenius
is charged to look into it and produces a report to
the Academy, in which he, on general grounds, argues
against a second Prize to Rutherford. 
His report includes the following statements (my translation):

\vspace{.2cm}
$\bullet$
There is very little sympathy for giving the same person two
Nobel Prizes.

$\bullet$
None of Rutherford's countrymen have nominated him for the Prize.

$\bullet$
Sir Ernest's meritorious contributions are so great and widely
known that his standing, and possibilities to do research would 
hardly be affected by a second Prize.

$\bullet$
He already occupies the highest position 
in the British Empire.
 
\vspace{.2cm}
For the 1924 Prize, Rutherford receives a nomination by 
David S. Jordan from Leland Stanford Jr. University.
Retired by then, Jordan had been a professor of natural 
sciences, an ichthyologist, and
the first president of Stanford University. He submitted
a few more nominations after 1924 but did not repeat his
nomination of Rutherford. In his 1924 nomination, he writes:

\vspace{.2cm}
"..the work following on his previous discovery of the
nuclear character of positive electricity, is a most
remarkable and extremely important line of research,
and taken with the high degree of excellence of all his
later work, makes him appear to me as a most suitable
candidate for the prize."

\vspace{.2cm}
Then there are no further nominations until 1931, 
when Johannes Stark (Nobel Laureate 1919) nominates Rutherford
for his work on alpha rays and atomic structure.
He writes (my translation)

\vspace{.2cm}
"Gentlemen, I am afraid you might be offended if I were
to justify to you in more detail the fundamental 
importance of the studies carried out by Rutherford. 
I consider it my scientific duty to inform you and
your Nobel Foundation 
that scientific justice and fairness require that you urgently 
award him the Nobel Prize in Physics."

\vspace{.2cm}
The response of the Committee to this nomination is
strange, to say the least. In 1931 the Committee, in
its report to the Academy, writes:

\vspace{.2cm}
"With all due respect for the importance of Rutherford's work,
the Committee is of the opinion
that these lie so close to the work for which he has been
given the Chemistry Prize that the awarding of a further
Prize is not justified." 

\vspace{.2cm}
Stark repeated his nomination four times
(1932, 1933, 1935 and 1937), i.e., until Rutherford passed away.

\vspace{.2cm}
Was Rutherford disappointed for not getting a second Prize?
We don't know but 
I don't believe so, as I will explain further down.  

\vspace{.1cm}
The case of Marie Curie
is different. I will not go into it in any detail
but would like to remind you that
in 1903 the Prize was divided into two halves. One half
went to Becquerel, "for his discovery of spontaneous
radioactivity" while the other half was further divided between
Pierre and Marie Curie, "... for their joint researches on
the radiation phenomena discovered by Professor Henri
Becquerel". Pierre Curie died in 1906. The Chemistry Committee
felt that Marie Curie's Prize in Physics did not give her the
recognition she deserved. It was she who
had discovered the chemical elements radium and 
polonium and, therefore, deserved
the 1911 Nobel Prize in Chemistry. Marie Curie had a total of three
nominations in physics and two in chemistry.

\vspace{.1cm}
There has never been a case
where a person all alone has received the total of 
two prizes, either in physics or in chemistry or
one in each discipline.

\section{Rutherford on the Photon, the Atomic Energy, and the Laws of Nature}

\subsection{The Photon}
In a letter dated January 26, 1917, Rutherford 
nominates Charles Barkla (1877 - 1944)
to Nobel Prize in Physics. Arriving late, this nomination
is taken to be valid for 1918. This is the only nomination
Barkla ever receives and is sufficient to earn him, in 1918, the
Nobel Prize for 1917, which had not yet been awarded. 
Barkla's "competitors", on the Nobel scene, Einstein
and Planck have each six nominations in 1918 and 
in 1919 Planck receives the 1918 Prize in Physics.
Rutherford nominates Barkla for

\vspace{.2cm}
"his important original contributions to our knowledge of 
the nature of X-rays, and particularly for his discovery
of the characteristic X-radiations of the
elements.

The proof that each element under certain conditions
emits an X-radiation of the element is a contribution that,
in my opinion, ranks only second in importance to the subsequent
discovery of the diffraction of X-rays by Laue. ..."

\vspace{.2cm}
\noindent
(Max von Laue had already been awarded the 1914 Nobel Prize.) 

\vspace{.1cm}
In praising Barkla, at the end of his nomination, 
Rutherford adds a surprising statement: 

\vspace{.2cm}
"Professor Barkla was throughout a staunch adherent of the
view that X-rays were a type of wave motion, and
championed this with vigour when a more materialistic
hypothesis appeared to be gaining ground."

\vspace{.2cm}
This means that in 1917 neither Barkla nor Rutherford believe in
Einstein's photon of 1905! However, a nomination letter
by Rutherford in 1929, proposing Richardson to the Prize,
suggests that he (perhaps) accepts the photon.

\subsection{The Atomic Energy}
Rutherford expresses his opinion on the future use of
atomic energy in a talk in 1933 by stating:

\vspace{.2cm}
"The transformation of the atom are of extraordinary interest
to scientists but we cannot control atomic energy to an
extent which would be of any value commercially, and I believe
we are not likely ever to be able to do so."

\vspace{.2cm}
How fortunate he was not to know about atomic bombs!

\subsection{The Laws of Nature}

Rutherford was not so keen
on quantum mechanics of Heisenberg 
and Schr\"odinger.
On uncertainty relations he writes in 1933
(the year in which Heisenberg was awarded the 1932 Prize
and Schr\"odinger and Dirac the 1933 Prize):

\vspace{.2cm}
"While the theory of indeterminacy is of great
theoretical interest as showing the limitations of
the present wave-theory of matter, its importance
in physics seems to me to have been much exaggerated
by many writers. It seems to me unscientific and also
dangerous to draw far-flung deduction from a 
theoretical conception which is incapable of 
experimental verification, either directly or indirectly."

\vspace{.2cm}
Being a truly great scientist, he would have, of course, 
in due time accepted the experimental evidence.
Rutherford was as he puts it: 

\vspace{.2cm}
"much amused at various
articles ... by writers ... who hold up their hands at the 
audacity of experimentations .. and sagely reflect how
Newton would have sat down and worked out the whole 
subject and then given a theory. It never occurs to them
that it would have wanted half a dozen Newtons to
accomplish the experimental work in a lifetime and even
they could not have put forward any more plausible theory
than we work on today. These dam'd fools ..."

\vspace{.2cm}
On the nature of our science he writes:

\vspace{.2cm}
"There is an error far too prevalent to-day that Science
progresses by the demolition of former well-established theories.
Such is very rarely the case. For example, it is often stated
that Einstein's general theory of relativity has overthrown
the work of Newton on gravitation. No statement can be
further from the truth. Their works, in fact, are
hardly comparable for they deal with different fields of
thought. So far as the work of Einstein is relative to
that of Newton, it is simply a generalisation and broadening
of its basis, in fact a typical case of mathematical and
physical development. In general a great principle is not
discarded, but is so modified that it rests on a broader
and more stable basis".

\vspace{.2cm}
Here Rutherford is making a very important point which
even today (i.e., 85 years later) many
people don't seem to understand. They don't trust science because
they believe that scientists keep on changing their opinions
on what is "true" - what was true yesterday is no longer
true today. They say physicists have shown that 
Newton was wrong. Surely, in the future, Einstein's picture of
the world will also be dumped into the dustbin of history. 

\vspace{.1cm}
It was typical of Rutherford that "he kept his feet firmly
on the ground and avoided the more speculative aspects of
physics".

\vspace{.1cm} 
I guess he would not have cared at all about the "Theory of
Everything" and such, by their nature, untestable hypotheses.
He might have accepted it
as a humorous concept, a naming used in the search
for acquiring new knowledge.
After all, he had a great sense of humour. 

\vspace{.1cm}
In conclusion
it seems that nobody is perfect but I venture
to say that "our dear Lord" was as perfect as anyone is allowed
to be, by the laws of nature.
For me, a more perfect person is hard to imagine!

\section{Rutherford and His Celebrity}

Rutherford is remarkable. He seems to violate a conjecture of
mine that reads:
\begin{equation}
C \vert C > = \vert 0>
\end{equation}
This equation, expressed in words, reads:
Celebrity operator $C$ acting on the state of creativity $\vert C>$ 
gives vacuum. In other words, Celebrity annihilates Creativity. 
Indeed, Celebrity absorbs such an enormous amount of time and
energy from its victims that there should be no room left for creativity.
Rutherford is aware of this problem.

\vspace{.1cm}
He becomes a mini-Celebrity soon after arriving in 
Cambridge in 1895. He
is asked to give demonstrations, talks, and is often
invited to dinners and expected to be social and entertaining.
Later on, during his period at McGill, he develops into
a real Celebrity. He is the "lion of the season" and the newspapers
are becoming radioactive. He is often the guest of honour and as
he puts it in a letter to his wife "It is not altogether
pleasant to be talked at, for four solid hours in
succession".

\vspace{.1cm}
He has much less time and
life is much tougher. He even has to cancel his private trips to 
have time to do some work. His letters give ample evidence
on this, such as:

\vspace{.2cm}
"It is important I should write it up as they 
are all following my trail, and if I am to have a chance
for a Nobel Prize in the next few years I must keep my
work moving."

\vspace{.2cm} \noindent
Then there is the Nobel Prize and he notes:

\vspace{.1cm}
"My correspondence alarms me by its dimensions."

\vspace{.1cm}
\noindent
Normally, in such a state, one would have no time
to do research. But he, somehow manages, not only to go on 
but to make outstanding discoveries.
Therefore, in his case, the above conjecture is not 
quite right and perhaps needs to 
be "supersymmetrized", like everything else in our field.
In other words, it is not Celebrity but Super-Celebrity operator that
annihilates creativity.

\vspace{.1cm}
Soon he becomes a Super-Celebrity and has even less time.
He has discovered the nucleus and proposed a new atomic
model.
In a letter to his Mom in 1912 Rutherford hints at the ensuing
events:

\vspace{.2cm}
"The last month has been filled with congresses and 
celebrations and I am glad they are now over and I
can settle down to three weeks' uninterrupted work
before the vacation."

\vspace{.2cm}
It is hard to imagine that he gets those three uninterrupted weeks. 
Nonetheless, he manages to make new discoveries, such as the
proton.

\vspace{.1cm}
It may amuse you to know that when Rutherford nominates Soddy
to the 1918 Nobel Prize in Chemistry he sends the
nomination to the Physics Committee! And his handwriting shows
that he has been in a hurry.
 
\vspace{.1cm}
Rutherford, a Super-Celebrity when he moves
from Manchester to Cambridge in 1919, turns very soon 
into a Hyper-Celebrity. Does Hyper-Celebrity
definitely kill Creativity?
 
\vspace{.1cm}
Rutherford is now under extreme external forces that cost
him a great deal of time.
He is "everywhere". He is the President of the 
Royal Society 1925-1930, a Baron, President of the Institute
of Physics 1931-1933, etc. 
Nonetheless he keeps on
working and making new discoveries up until he dies.
Eve recounts: 

\vspace{.2cm}
"On the occasion of one of his discoveries,
I said to him: "You are a lucky man, Rutherford, 
always on the crest of the wave!" To which he laughingly 
replied, "Well! I made the wave, didn't I?" and added soberly,
"At least to some extent.""

\vspace{.2cm}
Perhaps that's the explanation?

\section{Final Remarks}

Rutherford accomplished a great deal, among them:

$\circ$ \noindent He identified the $\alpha$ particles.

$\circ$ \noindent
He explained the origin of radioactivity with an
extremely bold idea - the transmutation of elements -
and gave us the exponential decay law and the concept
of half-lives.

$\circ$ \noindent He discovered the atomic nucleus.

$\circ$ \noindent He discovered and named the proton.

$\circ$ \noindent He predicted the existence of 
neutron and looked for it. The neutron was 
later discovered by one of his research students, James Chadwick.

$\circ$ \noindent He discovered tritium and helium-3, 
together with his research students.

$\circ$ \noindent He was great at building appartus and
detectors as well as doing the required theory. 

\vspace{.2cm}
The Empire, in recognition of his services, bestowed upon him
Knighthood (1914), the Order of Merit (1925), 
made him a Baron (1931) and interred
his ashes in Westminster Abbey (1937). 

\vspace{.1cm}
Rutherford was a generous person who gave a great deal of
credit to his collaborators, such as Chadwick and Soddy,
as well as many other people. His nominations testify
that he played down his own role. Those who knew him
seem to have really "loved" him. His research fellows 
admired him and several of them rose to great heights
in the society, for example Sir Ernest Marsden (1889-1970),
in New Zealand and Sir Mark Oliphant (1901-2000)
in Australia.
They were all very grateful to him.

\vspace{.1cm}
I guess if he would have wanted a second Nobel Prize he could
have given a slight hint to his distinguished 
colleagues and many of his people would have 
gladly nominated him. Only one person (John Cox) from 
the British Empire nominated him to his first Nobel Prize 
and no one for a second Prize.

\vspace{.1cm}
I believe that Rutherford would have loved 
to be here at this Conference in
Christchurch. As you may have noticed many talks were
concerned with one of his domains of expertise, radioactivity.
We saw a lot of $\alpha$'s. He would have also enjoyed
the talks on geophysical aspects of our science.

\vspace{.2cm}
What if we would have asked him for his advice? Perhaps,
to us theorist, he would have repeated one of his statements:

\vspace{.2cm}
"Spend more time in thinking and less in doing."

\vspace{.2cm}
And addressing some of you experimentalist, who 
have to deal with many co-workers and big budgets,
he might have added another one of his statements:

\vspace{.2cm}
"It is essential for you to take interest in the
administration of your own affairs or else the
professional civil servants would stop in .. and then
the Lord help you."

\section{Acknowledgements}

I wish to thank Stephen Parke and Francis Halzen for inviting
me to present this talk at Neutrino 2008.

My most sincere thanks go to Karl Grandin, Maria Asp and
Anne Miche de Malleray, at the Center for History of
Science, Royal Swedish Academy of Sciences,
Stockholm, for their kind
and cheerful reception whenever 
I have visited them to consult original material related 
to Nobel Prizes.

I am indebted to Arthur Stewart Eve (1862-1941) for his book
about Rutherford. I read it when I was a PhD student
and re-read it when preparing this talk. It is wonderful.

\appendix 
\section {The Sources of the Presented Material}

\vspace{.1cm}
Many years ago I became curious about why Rutherford didn't
get a second Nobel Prize. 
I started "digging" in the Nobel Archives in Stockholm.
The materials related to the Nobel, that I have presented
in this article, come from the Nobel Archives at "Center
for History of Science, Royal Swedish Academy of Sciences,
Stockholm". These Archives contain the annual reports that the Nobel
Committees submit to the Royal Swedish Academy of Sciences 
(in this article often referred to as the Academy). The reports
summarize the current status related to Nobel Prizes. They contain
information on who are the nominated candidates and what
they have been nominated for as well as the opinion
of the Committee. 

\vspace{.1cm}
One should keep in mind that the Committees are
expected to propose the candidates to be awarded,
but the decision is taken at Academy's plenum session where
all the members are invited to take part. They
may express their opinions and if they so wish
choose someone not suggested by the Committee.
This has happened several times.
There are no minutes of Nobel deliberations at Academy's
plenums.  

\vspace{.1cm}
Furthermore, the archives contain letters written by
members of the Academy who wish to state their (often conflicting) 
opinions, in order to make it known to their fellow academicians
and the posterity.

\vspace{.1cm}
Here, whenever I quote from the Committee reports
and letters by the members of the Academy, 
I am giving my own simple translation but I try to convey
correctly the sense of the original material, which is all
in Swedish. 

\vspace{.1cm}
In addition, the Nobel Archives contain the original nominations
and related correspondences,
such as hand-written letters by Rutherford, Einstein and many
other great scientists. It has been
a great pleasure for me to hold such letters in my hand
and read them. 

\vspace{.1cm}
In the case of Rutherford, most of the nominations are
in German. Here, I have given my own (again simple)
translation.

\vspace{.2cm}
I also present a number of extracts
from letters written by or to Rutherford taken 
from a wonderful book \cite{eve} by
Arthur Stewart Eve (1862-1941). The book was published in 1939, 
i.e., shortly after Rutherford's death. It bears
the title "Rutherford" and the subtitle "Being
the Life and Letters of the Rt Hon. Lord Rutherford,
O.M.". Here Rt Hon. means Right Honourable and O.M. stands 
for Order of Merit, an exclusive British award given by
the King/Queen. Eve himself
was a distinguished scientist ornamented with a string of
honours. He had known Rutherford for 35 years. 
They had been colleagues at McGill University, Canada,
and had become friends. Eve had the great privilege of having 
access to Rutherford's archive, put at
his disposal by Lady Rutherford.
I will assume that the letters have been correctly
typed. 

\vspace{.1cm}
Eve's book seems to have been written in a hurry.
I have found quite a
few errors (incorrect dates, typos, and misspelled names)
which could have easily been corrected in a later edition.
Alas there was no time for that! 
Eve died soon after the publication of his book. 
Nonetheless, in my opinion, it is a wonderful book.

\section {Some of the Actors in Rutherford's
Nobel Drama}

$\circ$
Svante August Arrhenius (1859 - 1927): 1903 Nobel Laureate in Chemistry.
Member of the Nobel Committee for Physics, 1900-1927. A highly 
knowledgeable man. The Nobel Archives clearly show that his
opinion and judgement mattered a great deal when choosing the Nobel 
Laureates. 

\vspace{.1cm}
\noindent
$\circ$
Johannes Stark (1874-1957): 1919 Nobel Laureate in Physics.
Nominated Rutherford to the Physics Prize five times in 1930's,
with no success.

\vspace{.1cm}
\noindent
$\circ$
Henrik S\"oderbaum (born 1862): Member of the Nobel Committee for
Chemistry 1900-1933. Did most of the ground work in connection
with the 1908 Nobel Prize to Rutherford.  

\vspace{.1cm}
\noindent
$\circ$
The (Theodor) Svedberg (1884 - 1971): 1926 Nobel Laureate in Chemistry.
Member of the Nobel Committee for Chemistry (1925-1964).
Svedberg was a physical chemist and a member 
of the Physics Class of the Academy.

\end{document}